\documentclass[10pt, conference, letterpaper]{IEEEtran}
\usepackage{url}
\usepackage[utf8]{inputenc}
\usepackage{xcolor}
\usepackage{amsmath}
\usepackage{multirow}
\usepackage{amssymb}
\usepackage{enumitem}
\usepackage{bm}
\usepackage{graphicx}

\usepackage{gensymb}
\newcommand{\red}[1]{\begin{color}{red}#1\end{color}}

\usepackage[acronyms,nonumberlist,nopostdot,nomain,nogroupskip]{glossaries}
\usepackage{tablefootnote}
\usepackage{booktabs}
\usepackage{tabularx}
\usepackage{epsfig}
\usepackage[outdir=images_final/]{epstopdf}
\usepackage{tikz}
\usepackage{pgfplots}
\pgfplotsset{compat=newest} 
\pgfplotsset{plot coordinates/math parser=false} 
\newlength\fheight
\newlength\fwidth
\usetikzlibrary{plotmarks,patterns,decorations.pathreplacing,backgrounds,calc,arrows,arrows.meta,spy,matrix}
\usepgfplotslibrary{patchplots,groupplots}
\usepackage{tikzscale}
\usepackage{siunitx}

\usepackage{multirow}

\renewcommand{\figurename}{Fig.}
\usepackage[font=scriptsize]{subcaption}
\usepackage[font=footnotesize]{caption}

\usepackage{mathtools}

\newcommand{\MP}[1]{\color{red}{\textbf{MP says: #1 }}\color{black}}
\newcommand{\MG}[1]{{\color{red}MG: #1}}
\newcommand{\ml}[1]{{\color{red}ML: #1}}
\newcommand{\pt}[1]{{\color{red}PT: #1}}

\newcommand{\rev}[1]{{\color{black}#1}}

\usepackage{dblfloatfix}    
\usepackage{colortbl}

\newacronym{3gpp}{3GPP}{3rd Generation Partnership Project}
\newacronym{adc}{ADC}{Analog to Digital Converter}
\newacronym{5g}{5G}{5th generation}
\newacronym{6g}{6G}{6th generation}
\newacronym{aimd}{AIMD}{Additive Increase Multiplicative Decrease}
\newacronym{am}{AM}{Acknowledged Mode}
\newacronym{amc}{AMC}{Adaptive Modulation and Coding}
\newacronym{aqm}{AQM}{Active Queue Management}
\newacronym{awgn}{AWGN}{Additive White Gaussian Noise}
\newacronym{balia}{BALIA}{Balanced Link Adaptation}
\newacronym{bdp}{BDP}{Bandwidth-Delay Product}
\newacronym{qos}{QoS}{Quality of Service}
\newacronym{pqos}{PQoS}{predictive quality of service}
\newacronym{bf}{BF}{Beamforming}
\newacronym{cc}{CC}{Congestion Control}
\newacronym{cdf}{CDF}{Cumulative Distribution Function}
\newacronym{cn}{CN}{Core Network}
\newacronym{cqi}{CQI}{Channel Quality Information}
\newacronym{cp}{CP}{Control Plane}
\newacronym{csirs}{CSI-RS}{Channel State Information - Reference Signal}
\newacronym{dc}{DC}{Dual Connectivity}
\newacronym{dce}{DCE}{Direct Code Execution}
\newacronym{dci}{DCI}{Downlink Control Information}
\newacronym{dl}{DL}{downlink}
\newacronym{dmr}{DMR}{Deadline Miss Ratio}
\newacronym{dmrs}{DMRS}{DeModulation Reference Signal}
\newacronym{e2e}{E2E}{end-to-end}
\newacronym{ecn}{ECN}{Explicit Congestion Notification}
\newacronym{edf}{EDF}{Earliest Deadline First}
\newacronym{enb}{eNB}{evolved Node Base}
\newacronym{epc}{EPC}{Evolved Packet Core}
\newacronym{es}{ES}{Edge Server}
\newacronym{fdma}{FDMA}{Frequency Division Multiple Access}
\newacronym{fdd}{FDD}{Frequency Division Duplexing}
\newacronym[firstplural=Radio Access Technologies (RATs)]{rat}{RAT}{Radio Access Technology}
\newacronym{fs}{FS}{Fast Switching}
\newacronym{ftp}{FTP}{File Transfer Protocol}
\newacronym{gnb}{gNB}{Next Generation Node Base}
\newacronym{harq}{HARQ}{Hybrid Automatic Repeat reQuest}
\newacronym{hetnet}{HetNet}{Heterogeneous Network}
\newacronym{hh}{HH}{Hard Handover}
\newacronym{hol}{HOL}{Head-of-Line}
\newacronym{ia}{IA}{Initial Access}
\newacronym{ieee}{IEEE}{Institute of Electrical and Electronics Engineers}
\newacronym{imt}{IMT}{International Mobile Telecommunication}
\newacronym{iot}{IoT}{Internet of Things}
\newacronym{ldpc}{LDPC}{Low-Density Parity Check}
\newacronym{los}{LOS}{Line of Sight}
\newacronym{lte}{LTE}{Long Term Evolution}
\newacronym{m2m}{M2M}{Machine to Machine}
\newacronym{ml}{ML}{machine learning}
\newacronym{mac}{MAC}{Medium Access Control}
\newacronym{mc}{MC}{Multi-Connectivity}
\newacronym{mcs}{MCS}{Modulation and Coding Scheme}
\newacronym{mec}{MEC}{Mobile Edge Cloud}
\newacronym{mi}{MI}{Mutual Information}
\newacronym{mimo}{MIMO}{Multiple Input, Multiple Output}
\newacronym{mmwave}{mmWave}{millimeter wave}
\newacronym{mptcp}{MPTCP}{Multipath TCP}
\newacronym{mr}{MR}{Maximum Rate}
\newacronym{mss}{MSS}{Maximum Segment Size}
\newacronym{mtd}{MTD}{Machine-Type Device}
\newacronym{mtu}{MTU}{Maximum Transmission Unit}
\newacronym{nfv}{NFV}{Network Function Virtualization}
\newacronym{nlos}{NLOS}{Non-Line-of-Sight}
\newacronym{nlosv}{NLOSv}{Vehicle Non-Line-of-Sight}
\newacronym{nr}{NR}{New Radio}
\newacronym{ofdm}{OFDM}{Orthogonal Frequency Division Multiplexing}
\newacronym{pdcch}{PDCCH}{Physical Downlink Control Channel}
\newacronym{pdcp}{PDCP}{Packet Data Convergence Protocol}
\newacronym{pdsch}{PDSCH}{Physical Downlink Shared Channel}
\newacronym{pdu}{PDU}{Packet Data Unit}
\newacronym{pf}{PF}{Proportional Fair}
\newacronym{pgw}{PGW}{Packet Gateway}
\newacronym{phy}{PHY}{Physical}
\newacronym{pbch}{PBCH}{Physical Broadcast Channel}
\newacronym[plural=\gls{mme}s,firstplural=Mobility Management Entities (MMEs)]{mme}{MME}{Mobility Management Entity}
\newacronym{prb}{PRB}{Physical Resource Block}
\newacronym{pss}{PSS}{Primary Synchronization Signal}
\newacronym{pscch}{PSCCH}{Physical Sidelink Control Channel}
\newacronym{pucch}{PUCCH}{Physical Uplink Control Channel}
\newacronym{pusch}{PUSCH}{Physical Uplink Shared Channel}
\newacronym{rach}{RACH}{Random Access Channel}
\newacronym{ran}{RAN}{Radio Access Network}
\newacronym{red}{RED}{Random Early Detection}
\newacronym{rf}{RF}{Radio Frequency}
\newacronym{rlc}{RLC}{Radio Link Control}
\newacronym{rlf}{RLF}{Radio Link Failure}
\newacronym{rrc}{RRC}{Radio Resource Control}
\newacronym{rrm}{RRM}{Radio Resource Management}
\newacronym{rr}{RR}{Round Robin}
\newacronym{rs}{RS}{Remote Server}
\newacronym{rsrp}{RSRP}{Reference Signal Received Power}
\newacronym{rss}{RSS}{Received Signal Strength}
\newacronym{rtt}{RTT}{Round Trip Time}
\newacronym{rw}{RW}{Receive Window}
\newacronym{rx}{RX}{Receiver}
\newacronym{sa}{SA}{standalone}
\newacronym{sack}{SACK}{Selective Acknowledgment}
\newacronym{sap}{SAP}{Service Access Point}
\newacronym{sc}{SC}{Single Carrier}
\newacronym{sch}{SCH}{Secondary Cell Handover}
\newacronym{scoot}{SCOOT}{Split Cycle Offset Optimization Technique}
\newacronym{sdma}{SDMA}{Spatial Division Multiple Access}
\newacronym{sinr}{SINR}{Signal to Interference plus Noise Ratio}
\newacronym{sl}{SL}{Sidelink}
\newacronym{sm}{SM}{Saturation Mode}
\newacronym{snr}{SNR}{Signal-to-Noise-Ratio}
\newacronym{son}{SON}{Self-Organizing Network}
\newacronym{ss}{SS}{Synchronization Signal}
\newacronym{srs}{SRS}{Sounding Reference Signal}
\newacronym{sss}{SSS}{Secondary Synchronization Signal}
\newacronym{tb}{TB}{Transport Block}
\newacronym{tcp}{TCP}{Transmission Control Protocol}
\newacronym{tdd}{TDD}{Time Division Duplexing}
\newacronym{tdma}{TDMA}{Time Division Multiple Access}
\newacronym{tfl}{TfL}{Transport for London}
\newacronym{tm}{TM}{Transparent Mode}
\newacronym{trp}{TRP}{Transmitter Receiver Pair}
\newacronym{tti}{TTI}{Transmission Time Interval}
\newacronym{ttt}{TTT}{Time-to-Trigger}
\newacronym{tx}{TX}{Transmitter}
\newacronym{ue}{UE}{User Equipment}
\newacronym{ul}{UL}{uplink}
\newacronym{uml}{UML}{Unified Modeling Language}
\newacronym{um}{UM}{Unacknowledged Mode}
\newacronym{utc}{UTC}{Urban Traffic Control}
\newacronym{vm}{VM}{Virtual Machine}
\newacronym{rsrq}{RSRQ}{Reference Signal Received Quality}
\newacronym{rssi}{RSSI}{Received Signal Strength Indicator}
\newacronym{crs}{CRS}{Cell Reference Signal}
\newacronym{nsa}{NSA}{Non Stand Alone}
\newacronym{mrdc}{MR-DC}{Multi \gls{rat} \gls{dc}}
\newacronym{endc}{EN-DC}{E-UTRAN-\gls{nr} \gls{dc}}
\newacronym{5gc}{5GC}{5G Core}
\newacronym{si}{SI}{Study Item}
\newacronym{iab}{IAB}{Integrated Access and Backhaul}
\newacronym{wf}{WF}{Wired-first}
\newacronym{hqf}{HQF}{Highest-quality-first}
\newacronym{pa}{PA}{Position-aware}
\newacronym{mlr}{MLR}{Maximum-local-rate}
\newacronym{wbf}{WBF}{Wired Bias Function}
\newacronym{mib}{MIB}{Master Information Block}
\newacronym{sib}{SIB}{Secondary Information Block}
\newacronym{rnti}{RNTI}{Radio Network Temporary Identifier}
\newacronym{dft}{DFT}{Discrete Fourier Transform}
\newacronym{kpi}{KPI}{Key Performance Indicator}
\newacronym{ppp}{PPP}{Poisson Point Process}
\newacronym{v2v}{V2V}{Vehicle-to-Vehicle}
\newacronym{wave}{WAVE}{Wireless Access in Vehicular Environments}
\newacronym{udp}{UDP}{User Datagram Protocol}
\newacronym{upa}{UPA}{Uniform Planar Array}
\newacronym{fec}{FEC}{Forward Error Correction}
\newacronym{v2x}{V2X}{Vehicle-To-Everything}
\newacronym{psfch}{PSFCH}{Physical Sidelink Feedback Channel}
\newacronym{pssch}{PSSCH}{Physical Sidelink Shared Channel}
\newacronym{csma}{CSMA}{Carrier Sense Multiple Access}
\newacronym{v2n}{V2N}{Vehicle-to-Network}
\newacronym{wlan}{WLAN}{Wireless Local Area Network}
\newacronym{cav}{CAV}{Connected and Autonomous Vehicle}
\newacronym{v2i}{V2I}{Vehicle-to-Infrastructure}
\newacronym{d2d}{D2D}{Device-to-Device}
\newacronym{c-its}{C-ITS}{Connected Intelligent Transportation System}
\newacronym{fr2}{FR2}{Frequency Range 2}
\newacronym{fr1}{FR1}{Frequency Range 1}
\newacronym{bs}{BS}{Base Station}
\newacronym{sdu}{SDU}{Service Data Unit}
\newacronym{csi}{CSI}{Channel State Information}
\newacronym{scs}{SCS}{Subcarrier Spacing}
\newacronym{sumo}{SUMO}{Simulation of Urban MObility}
\newacronym{prr}{PRR}{Packet Reception Ratio}
\newacronym{edca}{EDCA}{Enhanced Distribution Channel Access}
\newacronym{sdap}{SDAP}{Service Data Adaptation Protocol}
\newacronym{iiot}{IIoT}{Industrial Internet of Things}
\newacronym{agv}{AGV}{Automated Guided Vehicles}
\newacronym{cm}{C/M}{Controller/Master}
\newacronym{soa}{SoA}{State-of-the-Art}
\newacronym{snpn}{SNPN}{Standalone Non-Public Network}
\newacronym{pninpn}{PNI-NPN}{Public Network Interface Non-Public Network}
\newacronym{urllc}{URLLC}{Ultra-Reliable Low-Latency Communication}
\newacronym{embb}{eMBB}{enhanced Mobile BroadBand}
\newacronym{ai}{AI}{Artificial Intelligence}
\newacronym{mab}{MAB}{Multi-Armed Bandit}
\newacronym{su}{SU}{Scheduling Unit}
\newacronym{ra}{RA}{Random Agent}
\newacronym{na}{NA}{Neural Agent}
\newacronym{ucba}{UCB-A}{UCB Agent}
\newacronym{tsa}{TS-A}{Thompson Sampling Agent}
\newacronym{ucb}{UCB}{Upper Confidence Bound}
\newacronym{ts}{TS}{Thompson Sampling}
\newacronym{inf}{InF}{Indoor Factory}
\newacronym{infsl}{InF-SL}{Indoor Factory - Sparse Clutter, Low BS}
\newacronym{infdl}{InF-DL}{Indoor Factory - Dense Clutter, Low BS}
\newacronym{infsh}{InF-SH}{Indoor Factory - Sparse Clutter, High BS}
\newacronym{infdh}{InF-DH}{Indoor Factory - Dense Clutter, High BS}
\newacronym{us}{US}{Uplink Scheduler}
\newacronym{nn}{NN}{Neural Network}
\newacronym{das}{DAS}{Distributed Antenna System}
\newacronym{rb}{RB}{Resource Block}
\newacronym{rl}{RL}{Reinforcement Learning}
\newacronym{uav}{UAV}{Unmanned Aerial Vehicle}
\newacronym{5gacia}{5G-ACIA}{5G Alliance for Connected Industries and Automation}
\newacronym{drl}{DRL}{Deep Reinforcement Learning}

\def\si{\tikz\fill[scale=0.4](0,.35) -- (.25,0) -- (1,.7) -- (.25,.15) -- cycle;}

\definecolor{steelblue}{RGB}{176,196,222}

\usepackage{hyperref}
\usepackage[capitalize]{cleveref}

\makeglossaries
\begin{document}

	
	\title{Distributed Resource Allocation for URLLC in \\ IIoT Scenarios: A Multi-Armed Bandit Approach}
	
	\author{\IEEEauthorblockN{Francesco Pase$^{\star }$, Marco Giordani$^{\star }$, Giampaolo Cuozzo$^{\circ }$, Sara Cavallero$^{\circ }$, \\ Joseph Eichinger$^{\dagger }$, Roberto Verdone$^{\circ }$, Michele Zorzi$^{\star }$\medskip}
		\IEEEauthorblockA{
			$^{\star}$WiLab and University of Padova, Italy. Email: \texttt{\{name.surname\}@dei.unipd.it}\\
			$^{\circ}$WiLab and University of Bologna, Italy. Email: \texttt{\{name.surname\}@unibo.it}\\
			$^{\dagger}$Huawei Technologies, Munich Research Center, Germany. Email: \texttt{joseph.eichinger@huawei.com}}}

	\maketitle

	\begin{abstract}
		This paper addresses the problem of enabling inter-machine \gls{urllc} in future 6G \gls{iiot} networks.
		As far as the \gls{ran} is concerned, centralized pre-configured resource allocation requires scheduling grants to be disseminated to the \glspl{ue} before uplink transmissions, which is not efficient for URLLC, especially in case of flexible/unpredictable traffic. To alleviate this burden, we study a distributed, user-centric scheme based on machine learning in which \glspl{ue} autonomously select their uplink  radio resources without the need to wait for scheduling grants or preconfiguration of connections. Using simulation, we demonstrate that a \gls{mab} approach represents a desirable solution to allocate resources with URLLC in mind in an IIoT environment, in case of both periodic and aperiodic traffic, even considering highly populated networks and aggressive traffic. 
	\end{abstract}
			\begin{tikzpicture}[remember picture,overlay]
		\node[anchor=north,yshift=-10pt] at (current page.north) {\parbox{\dimexpr\textwidth-\fboxsep-\fboxrule\relax}{
				\centering\footnotesize This paper has been accepted for presentation at the 2022 IEEE Globecom Workshops (GC Wkshps): Future of Wireless Access and Sensing for Industrial IoT (FutureIIoT). \textcopyright 2022 IEEE. \\
				Please cite it as: F. Pase, M. Giordani, G. Cuozzo, S. Cavallero, J. Eichinger, R. Verdone, M. Zorzi, “Distributed Resource Allocation for URLLC in IIoT Scenarios: A Multi-Armed Bandit Approach,” IEEE Globecom Workshops (GC Wkshps): Future of Wireless Access and Sensing for Industrial IoT (FutureIIoT), Rio de Janeiro, Brazil, 2022.\\
				}};
	\end{tikzpicture}
	\glsresetall
	
	\section{Introduction}
	\label{sec:intro}
	With early \gls{5g} deployments already rolled out, the research community is discussing use cases, requirements, and enabling technologies towards \gls{6g} systems~\cite{giordani2020toward}. Among other services, 6G will introduce new communication interfaces and innovative architectures to support the {\gls{iiot}} in 2030 and beyond, where the 6G network connects sensors and machines in factories, plants, mines, to enable analytics, diagnostics, monitoring, asset tracking, as well as process, regulatory, supervisory, and safety control~\cite{lee2015cyber}. 
	In this context, the need for robots to complete cooperative operations that require high precision and coordination in real time comes with its own set of requirements, e.g., in terms of reliability (up to 99.99999\%) and latency (below 1 ms, or even 0.1 ms, in the radio part), making it crucial to support \gls{urllc} in the industrial domain~\cite{wollschlaeger2017future}.  The factory of the future will further operate to support high-density deployments  of machines and end users.
	
	In this context, the time introduced by the \gls{ran} operations, from routing and scheduling to resource allocation and  modulation, represents one of the most impactful latency components. Specifically, a \emph{centralized pre-configured} scheduling protocol usually requires the prior exchange of scheduling requests (grants) to (from) the \gls{gnb}, which is not compatible with \gls{urllc} in IIoT scenarios~\cite{cuozzo2022enabling, Yang2021}.
	To partially address this issue, 3GPP NR supports \emph{semi-persistent} and \emph{grant-free} communication in the \gls{ul}~\cite{3gpp.38.321}, in which the network pre-allocates radio resources, thereby eliminating the need for \glspl{ue} to wait for network grants before transmission.
	However, reserving resources to dedicated UEs can be inefficient if traffic demands are aperiodic~\cite{lucas2019capacity}, and it is not possible to anticipate when resources will be needed~\cite{boban2021predictive}.
	
	Another solution is to design a \emph{user-centric} architecture (as foreseen in 6G~\cite{wang2017machine}) in which end machines make autonomous decisions, ``disaggregated'' from the network~\cite{pase2020convergence}. Along these lines, in this paper we explore the feasibility of a decentralized/distributed scheduling algorithm that, exploiting \gls{ml} technologies, allows \glspl{ue} to optimize their \gls{ul} transmission strategies by autonomously selecting the available physical resources. 
	This framework is able to learn from the application, 
	and could work well even considering architectures for IIoT scenarios in which communication is on the sidelink, with no or limited support from the gNB~\cite{5gaciaiiot}. 
	
	Despite this potential, however, distributed scheduling may create collisions during communication, raising the question of whether this approach is compatible with URLLC applications.
	To this aim, we apply the \gls{mab} theory~\cite{intro_mab} to evaluate how autonomous machines should select transmission resources based on previous scheduling decisions and the effect they produced on the network in terms of reliability.
	While the \gls{mab} approach is well known, most related work focused on~\gls{dl}~\cite{liu2020data}, cellular~\cite{halabian2019distributed}, or IoT~\cite{hussain2020machine} networks. In turn, we consider a UL scenario modeled according to the ``Motion Control'' 5G-ACIA geometry (in which a~remote server sends commands to control the moving parts of machines), thus ensuring that our results are representative of a typical IIoT environment. 
	Other notable papers consider vehicular scenarios \cite{helin2019,Le2019}, where the target is to enable \gls{urllc} for vehicle-to-vehicle communications via \gls{drl}. However, we argue that for \gls{iiot} use cases, state-of-the-art \gls{mab} algorithms may better exploit the strong correlation typical of the industrial environment while, at the same time, reducing the computational complexity and training time to converge to optimal solutions, compared to more sophisticated \gls{drl} alternatives.
	
	We perform simulations with both periodic and aperiodic traffic, and as a function of the UEs' density and spatial distribution, the traffic periodicity (thereby modeling aggressive or conservative applications), and the transmit power, thus considering a low-power performance regime. From our results, we conclude that the Thompson Sampling agent~\cite{daniel_benji} is a promising candidate method to minimize the collision probability even in the presence of unscheduled transmissions.
	
	The rest of the paper is organized as follows. In Sec.~\ref{sec:problem_formulation_and_system_model} we present the distributed resource allocation problem, in Sec.~\ref{sec:mab_agents} we introduce possible ML methods based on MAB to solve it, and in Sec.~\ref{sec:perf} we describe our simulation setup and discuss our main results. Finally, Sec.~\ref{sec:conclusions_and_future_work} concludes our work with suggestions for future research.

	\section{Problem Formulation and System Model} 
	\label{sec:problem_formulation_and_system_model}
	
	\begin{figure}[t!]
		\setlength{\belowcaptionskip}{-0.7cm}
		\begin{center}
			\includegraphics[width=0.8\linewidth]{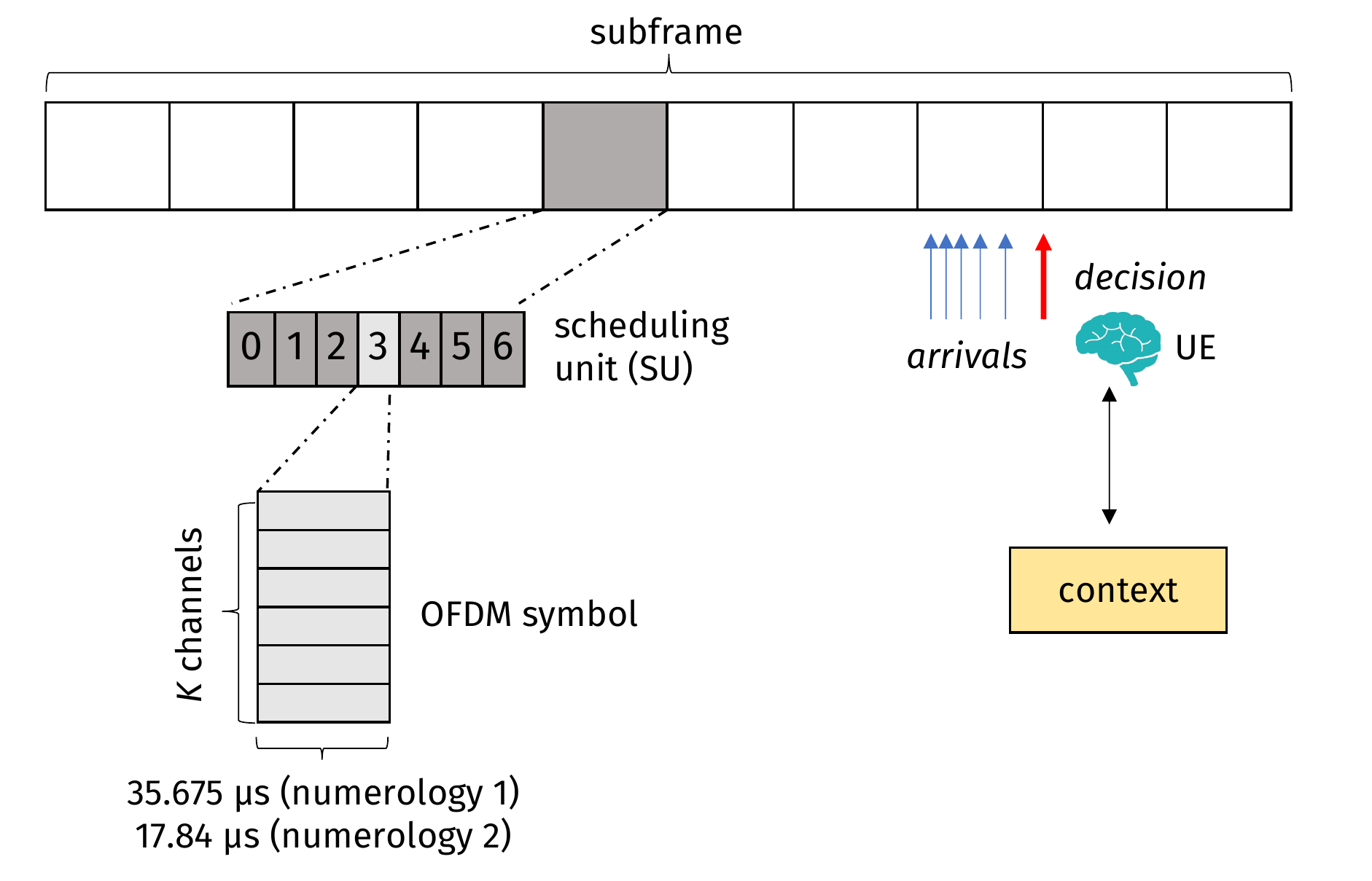}
			\caption{Transmission structure.} 
			\label{fig:intelligent_ue}
		\end{center}
	\end{figure}
	
	We consider an \gls{ofdm} system in which devices, also denoted as {agents} in machine learning parlance, are located in a factory environment, and have to autonomously choose the orthogonal channel to be used for UL transmissions. The time domain is discretized into intervals of duration equal to the \gls{ofdm} symbol (with a \gls{su} consisting of $7$ \gls{ofdm} symbols), whose duration depends on the adopted NR numerology. The frequency domain is also discretized into $K$ orthogonal channels, whose size depends on the available bandwidth $B$ and the subcarrier spacing $\Delta f$. 
	
	At the beginning of each \gls{su}, the agents make their scheduling decisions, that is the channel to be used for transmission, as shown in Fig.~\ref{fig:intelligent_ue}. 
	Unlike in a centralized pre-configured resource allocation approach, in which radio resources are scheduled by the \gls{gnb} via scheduling grants, we study the feasibility of a decentralized algorithm based on ML in which each agent autonomously optimizes its channel selection policy relying only on the \gls{gnb} feedback, without prior ad hoc message exchange with the \gls{gnb} itself. The rationale behind this scheme is to exploit the underlying correlations typical of the IIoT traffic to avoid the transmission of centralized scheduling grants, thus reducing the end-to-end latency and promoting \gls{urllc}.
	
	If multiple agents use the same physical channel during a specific \gls{su}, we assume that their packets are lost due to a \emph{collision} event. At the end of each \gls{su}, the \gls{gnb} broadcasts a message indicating in which channel(s) data were successfully received. This message is used by the pool of agents to optimize their subsequent decision strategies, and achieve coordination without communication.

	We formalize the problem using the \gls{mab} framework, which is used to model many sequential decision processes in computer science and engineering~\cite{intro_mab}. In this particular multi-agent scenario, there are $N$ agents, i.e., the $N$ \glspl{ue}, interacting with the same environment. Whenever an agent $n\in \{1, \dots, N\}$ generates a new packet during \gls{su} $t$, it schedules its transmission at the beginning of \gls{su} $t+1$, choosing one among the $K$ available channels, which will be used for transmission for the whole \gls{su} duration. According to the \gls{mab} notation, we refer to the action of using channel $k \in \mathcal{K} = \{1, \dots, K\}$ as ``playing the arm'' $k$.
	At the end of \gls{su} ${t+1}$, the message received from the \gls{gnb} is converted into a reward $r_{n,t}$, indicating whether or not the transmission was successful, i.e., $r_{n,t}= 1$ or $r_{n,t} = 0$, respectively: maximizing the reward implies transmitting the data successfully in low latency, as there is no need to exchange scheduling grants between the \glspl{ue} and the \gls{gnb}, leading to the \gls{urllc} objective. 
	In our model, we assume that the reward behind each action is sampled from a Bernoulli distribution with unknown parameter $\mu_n(k_{n,t})$, which depends on the action taken by the agent, and captures the probability of the other agents transmitting at the same time. Thus, in each \gls{su} $t$,  agent $n$ samples an action $k \in \mathcal{K}$ according to its policy $\pi_{n}: \mathcal{H}^{t-1} \rightarrow \Delta_K$, which is, in general, a map from history ${H_n(t-1) = \left\{ \left( k_{1,n}, r_{1, n} \right), \ldots,  \left( k_{t-1, n}, r_{t-1, n}\right)\right\} \in \mathcal{H}^{t-1}}$ to a probability distribution over the action set $\mathcal{K}$, where $\Delta_K$ denotes the $K$-simplex. The history vector $H_n(t-1)$ is used by the agent to optimize its policy $\pi_{n}$, so as to maximize the expected cumulative reward $R(\pi_n, T) = \mathbb{E}_{\pi_n} \left[ \sum_{t=1}^T \mu_n\left(k_{n,t}\right) \right]$.
	
	
	\section{\acrfull{mab} Agents} 
	\label{sec:mab_agents}
	
	To solve the problem in Sec.~\ref{sec:problem_formulation_and_system_model} and maximize the reward, many algorithms have been proposed in the literature over the past years~\cite{intro_mab}. 
	In this paper, we study the performance of different \gls{mab} agents to solve the problem of distributed resource allocation, in the specific context of \gls{urllc} for \gls{iiot}.

	\paragraph{\gls{ra}} It implements the simplest decision scheme, and is used as a lower bound. Nonetheless, it represents well the case of 5G NR grant-free scheduling, where the access decision is random, and re-transmissions are optimized to achieve reliability~\cite{Liu2021}. In particular, in each \gls{su}, the \gls{ra} selects uniformly, at random, one of the $K$ arms, and no learning is involved.

	\paragraph{\gls{ucba}} It implements the \gls{ucb} \rev{algorithm~\cite{lattimore_2020}}, i.e., the agent plays, in each \gls{su} $t$, the arm $k_t$ such that 
	\begin{equation}
	k_t = \text{argmax}_{k \in \mathcal{K}} \left[ Q_t(k) + c \sqrt{\frac{\log t}{n_t(k)}}\right],
	\label{eq:kt}
	\end{equation}
	where $Q_t(k)$ is the empirical average at step $t$ of the experienced rewards for arm $k$, $n_t(k)$ is the number of times arm $k$ has been played until time step $t$, and $c$ is an exploration parameter to be optimized. 
	In Eq.~\eqref{eq:kt}, $Q_t(k)$ represents the \textit{exploitation} part, as it is related to the past experience, while $\sqrt{{\log t}/{n_t(k)}}$ quantifies the uncertainty around the empirical average, and decreases as we collect more samples, i.e., as $n_t(k)$ increases. The larger this second term for an action $k$, i.e., the uncertainty of its performance, the higher the probability of choosing that arm, meaning that we need more samples to have a good estimate of its related reward. This principle is also known as \textit{``optimism in the face of uncertainty.''}

	\paragraph{\gls{tsa}} The agent adopts a \textit{Bayesian inference} approach to identify the most promising arms. In particular, \gls{tsa} builds a distribution for each reward, thus modeling not only its mean, but the whole \rev{statistics~\cite{daniel_benji}}. Given that our problem includes a binary reward $\{0,1\}$ behind each arm, it is quite natural to model the rewards according to a Bernoulli distribution, which is parameterized by the {success probability vector}~$\bm{\mu} = \left( \mu_1, \dots, \mu_K\right)$, where $\mu_k$ represents the average unknown reward behind arm $k \in \mathcal{K}$. 
	Following the Bayesian framework, parameter $\mu_k$ of arm $k$ is modeled as a $\text{Beta}(\alpha_k, \beta_k)$ random variable, where $\alpha_k$ counts the number of successful transmissions after playing arm $k$, and $\beta_k$ represents the number of collisions. Therefore, the mean of $\mu_k$ is equal to $\alpha_k/(\alpha_k + \beta_k)$. The Beta distribution parameters are initialized to $\{\alpha_k =1, \beta_k=1\}$ for all $k \in \{1, \dots, K \}$. 
	
	As the \gls{tsa} collects more data, $\alpha_k$ and $\beta_k$ are updated accordingly, inducing biased probabilities for the different arms. These informed distributions are also called \textit{posterior probabilities}, in Bayesian parlance. Whenever the agent makes a decision, i.e., it chooses a physical channel based on the probability of that channel not being accessed by other agents in that time interval, it samples a vector $\bm{\mu} = (\mu_1, \dots, \mu_K)$,
	and plays the arm $k^*$ such that $k^* = \text{argmax}_{k} \{\mu_k\}$. This algorithm is known as the \gls{ts} \rev{algorithm~\cite{daniel_benji}}

	\paragraph{\gls{na}} The \gls{na} is equipped with a small-size \gls{nn} used to represent its decision policy. In particular, the agent receives, as an input, context information $s_t \in \mathcal{S}$ from the environment, thus the problem is formulated as a contextual \gls{mab}, i.e., \rev{the average reward depends on the played arm $k_{n,t}$, and on the state $s_{n,t}$~\cite{lattimore_2020}}. The \gls{nn} input represents the feedback on the results of the last transmission attempt, broadcast by the \gls{gnb}. As such, the input data is a vector of $K+1$ entries: the first $K$ values are the results of the transmission attempts in the $K$ orthogonal channels, whereas the last value indicates whether it is a first-time transmission or a re-transmission. 
	Again, the $0/1$ reward given to failed/successful transmission, respectively, is used by the \gls{na} to optimize the \gls{nn} parameters, and maximize the given rewards. The model is an adaptation of that in~\cite{differentible_mab}.

	\emph{Remark.} The \gls{ucb} and \gls{ts} algorithms exhibit good theoretical properties in terms of convergence time to optimal strategies, as long as some critical assumptions are satisfied~\cite{lattimore_2020}:
	\begin{enumerate}
		\item The rewards behind each action need to exhibit a sub-Gaussian distribution. Any distribution with limited support has this property, which is also verified in our setting.
		\item The reward samples after playing action $k$ are i.i.d. This assumption is more critical in real scenarios, and in particular in our problem. In fact, each agent interacts with many other devices, and so the rewards depend on the actions of the other agents, which are continuously learning and changing their decision schemes. This leads to highly non-stationary environments, meaning that the reward distribution may change over time. However, empirical results show that state-of-the-art MAB algorithms can still be applied even though the \rev{stationarity} assumption for the rewards is not satisfied~\cite{Bonnefoi2018}.
	\end{enumerate}
	
	In Sec.~\ref{sub:numerical_results} we compare the performance of the MAB agents presented above, and provide guidelines towards the best schemes to satisfy URLLC requirements for~IIoT.
	
	\section{Performance Evaluation} 
	\label{sec:perf}
	In this section, after introducing our simulation setup, we evaluate the performance of the proposed distributed resource allocation scheme implementing one of the MAB agents described in Sec.~\ref{sec:mab_agents}, in  different IIoT scenarios.
	
	\subsection{Simulation Setup} 
	\label{sub:simulation_setup}
	
	
	
	End machines transmit at frequency $f_c=3.5$ GHz and with a bandwidth of $B=20$ MHz.
	The subcarrier spacing is set to $\Delta f = 30$ KHz (i.e., 3GPP NR numerology 1), which results in $K = 55$ orthogonal channels, and an \gls{ofdm} symbols duration of $T_{\rm OFDM} \simeq 35.675\,\mu$s~\cite{polese20193gpp}. With an \gls{su} of 7 OFDM symbols, we get an \gls{su} duration of $T_{SU} \simeq 0.25$ ms. We assume that, whenever a packet is to be sent, it can be transmitted within one \gls{su}. If two or more \glspl{ue} select the same \gls{ul} channel for transmission in the same \gls{su}, we consider those packets to be lost (due to a collision event). Assuming that the \gls{gnb} feedback (informing about the collision) is received within the current \gls{su}, the retransmission can be scheduled in the subsequent~\gls{su}.
	
	The factory floor is characterized according to the 5G-ACIA ``Motion Control'' scenario, as described in~\cite{5gaciaiiot}. Hence, the geometry is modeled as a parallelepiped of length $\ell=15$~m, width $w=15$~m, and height $h=3$~m, and machines are randomly and uniformly distributed inside the factory. The \gls{gnb} is located at the center of the ceiling, and communicates with power $P_{\rm TX, DL}=30$ dBm. The transmit power of the UEs is set to  $P_{\rm TX, UL}\in\{8,10,23\}$ dBm. 
	Also, we consider omnidirectional transmissions, therefore the antenna gain is fixed to $G=1$ for both the UEs and the gNB.
	The channel model is based on the 3GPP \gls{inf} scenario~\cite{3gpp.38.901}, where UEs are assumed to communicate in \gls{nlos} if the joining line between the UE's and the gNB's centers intersects one or more machines.
	
	In our simulations, the traffic can be either \emph{periodic} or \emph{quasi-periodic}.
	In the first case, packets are generated at constant periodicity $\tau$.
	In the second case, the application still generates packets with periodicity $\tau$, upon which a random component $t_{\rm off}$ of $\{ -2, -1, 0, +1, +2 \}$ \gls{ofdm} symbols is added. 
	
	The performance of the different MAB agents' policies is assessed in terms of \emph{successful transmission rate} $S_{TX}$, which indicates the ratio between the successfully received packets and the total number of attempts within one SU, averaged over 1\,000 steps, as a function of the traffic periodicity $\tau$, the number of UEs $N$, and the UL transmission power $P_{\rm TX, UL}$. 
	Notice that $S_{TX}$ is inversely proportional to the number of re-transmissions and, as such, represents well the theoretical rewards $r_{n,t}$ of the MAB agents.

	\subsection{Numerical Results} 
	\label{sub:numerical_results}
	
	\begin{figure}[t!]
		\centering
		\begin{subfigure}[b]{0.5\textwidth}
			\centering
			\includegraphics[width=0.8\linewidth]{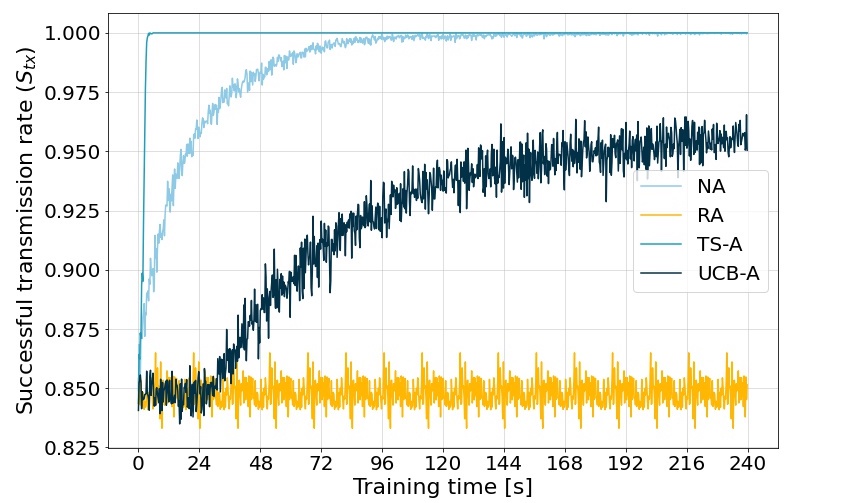}
			\caption{Periodic traffic.}
			\label{fig:acc_vs_time_periodic}
		\end{subfigure}
		\begin{subfigure}[b]{0.5\textwidth}
			\centering
			\includegraphics[width=0.8\linewidth]{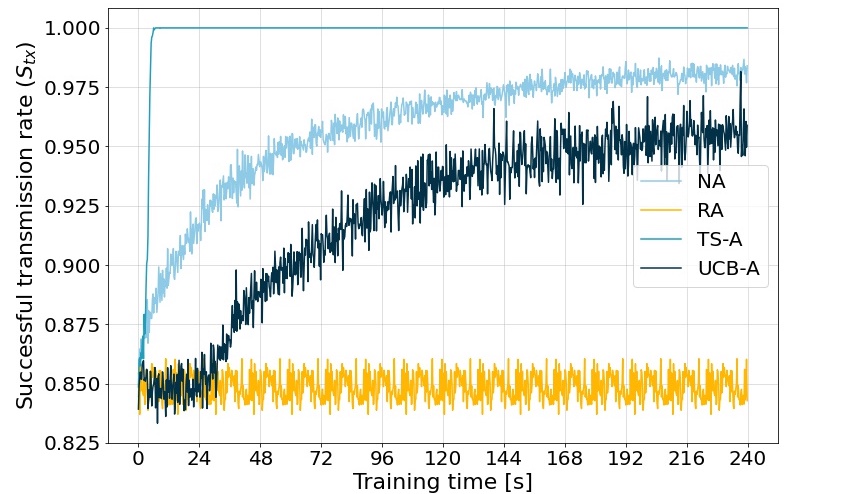}
			\caption{Quasi-periodic traffic}
			\label{fig:acc_vs_time_quasi_periodic}
		\end{subfigure}
		\caption{$S_{TX}$ vs. the training time, for different MAB agents, with periodic and quasi-periodic traffic, $\tau = 1.5$, and $N=50$.\vspace{-0.53cm}}
		\label{fig:training}
	\end{figure}

	\textbf{Impact of the training.} In Fig.~\ref{fig:training} we analyzed the training curve
	of the agents with periodic and quasi-periodic traffic, with a periodicity $\tau = 1.5$ ms, and considering $N=50$ UEs in the system, for a total training time of $T=240$~s. 
	For the periodic case, we observe from Fig.~\ref{fig:acc_vs_time_periodic} that \gls{tsa} is the best performing agent. In particular, the \gls{ts} agents are able to learn their optimal strategy, achieving zero collisions (i.e., $S_{TX}=1$, our target for URLLC) in a very short training time ($< 10$ s).
	\gls{na} achieves a similar performance to that of \gls{tsa}, though after a longer training process. This is due to the fact that
	\gls{na} needs more interactions with the system to optimize the network parameters, thus slowing down the training phase. 
	For \gls{ucba}, the exploration parameter $c$ in Eq.~\eqref{eq:kt} was set to 2, as it showed the most stable configurations in our experiments. Still, it results in an even slower convergence compared to \gls{na}, due to the fact that it struggles to achieve coordination. Also, \gls{ucba} presents significant oscillations over time, due to the impact of collisions and retransmissions.
	As expected, \gls{ra} (our baseline) performs poorly, and there is no improvement over time, as feedback signals are not exploited by the algorithm to adjust the access~scheme.
	
	For the quasi-periodic case, we observe from Fig.~\ref{fig:acc_vs_time_quasi_periodic} that \gls{tsa} presents again the best performance despite the more complex scenario, converging to zero collisions within $15$~s. Now, \gls{na} no longer achieves perfect convergence within the training time, suggesting that it cannot work well in non-stationary multi-agent scenarios, or deal with non-deterministic traffic requests. However, we believe that, with a better tuned training process, and with more relevant context information as input, the final performance would reasonably improve.
	Finally, \gls{ucba} and \gls{ra} perform similarly to the case of periodic traffic.
	\smallskip
	

	\begin{figure}[t!]
		\setlength{\belowcaptionskip}{-0.80cm}
		\begin{center}
			\includegraphics[width=0.85\linewidth]{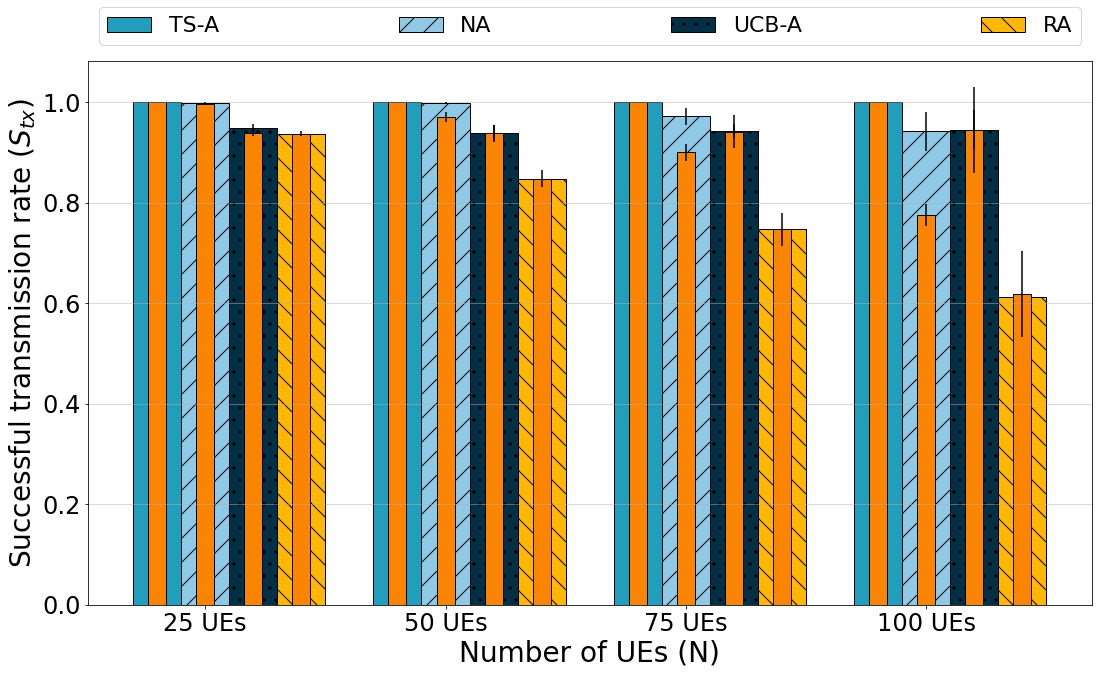}
			\caption{$S_{TX}$ $\pm$ one standard deviation vs. $N$ for different MAB agents, after a training time of 60 s, with $\tau = 1.5$ ms, with periodic (wide bars) and quasi-periodic (narrow bars) traffic.}
			\label{fig:accuracy_vs_users}
		\end{center}
	\end{figure}

	\textbf{Impact of the number of users.} 
	In Fig.~\ref{fig:accuracy_vs_users} we evaluate the performance of the MAB agents as a function of $N \in \{ 25, 50, 75, 100\}$. In particular, we studied the statistics of the successful transmission rate $S_{TX}$ after $60$ s of training, where again the total training time is set to $T=240$ s. 
	First, we observe that \gls{tsa} converges to the optimal scheme (i.e., $S_{TX}=1$) within $60$ s in all configurations, thus achieving coordination without communication even in dense ($N=100$) networks.
	Second, \gls{na} outperforms \gls{ucba} with periodic traffic, but suffers with quasi-periodic traffic: notably, $S_{TX}$ decreases by 10\% in the quasi-periodic case, for $N=100$. This is due to the fact that \gls{na} implements and exploits an \gls{nn} to optimize its decisions, thus the learning phase can take more time in the most complex scenarios.
	Interestingly, compared to other agents, \gls{ucba}'s performance is less sensitive to $N$, and eventually outperforms NA's approach in the most crowded scenarios. On the downside, it exhibits wider oscillations, i.e., higher standard deviation in Fig.~\ref{fig:accuracy_vs_users}, an indication of a less stable behavior of the agent in non-stationary environments.
	\smallskip
	
\begin{figure}[t!]
		\setlength{\belowcaptionskip}{-0.5cm}
		\begin{center}
			\includegraphics[width=0.83\linewidth]{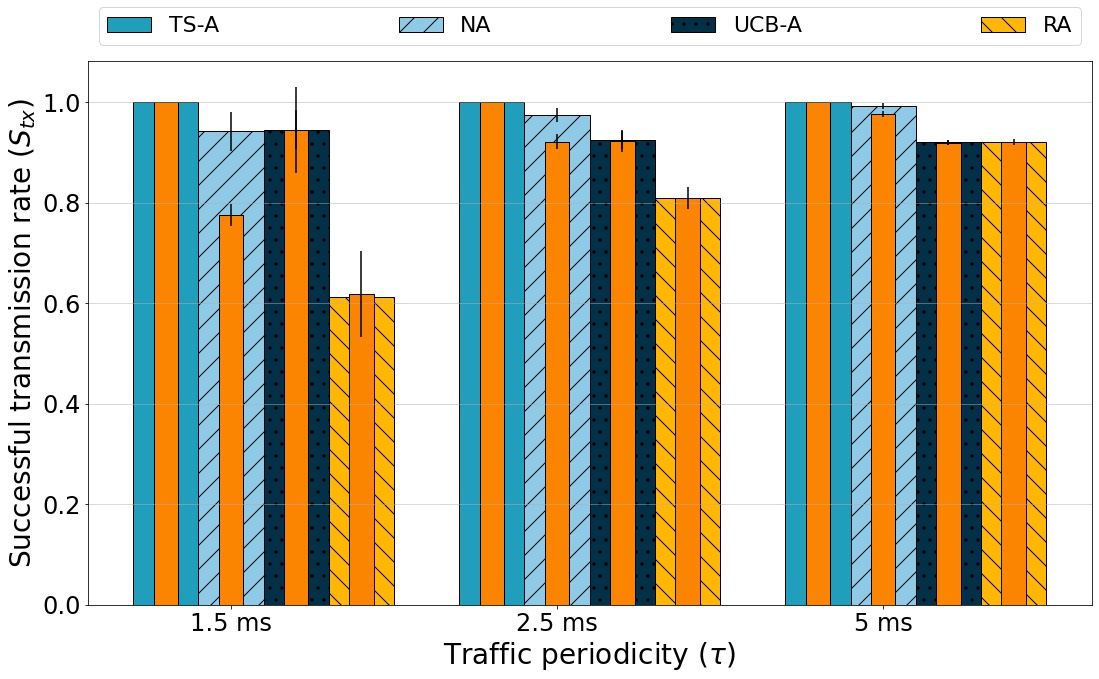}
			\caption{$S_{TX}$ $\pm$ one standard deviation vs. $\tau$ for different MAB agents, after a training time of 60 s, with $N = 100$, with periodic (wide bars) and quasi-periodic (narrow bars) traffic.}
			\label{fig:accuracy_vs_period}
		\end{center}
	\end{figure}

	\textbf{Impact of the traffic periodicity.} 
	Fig.~\ref{fig:accuracy_vs_period} explores the effect of the traffic periodicity $\tau$ on the successful transmission rate $S_{TX}$. As expected, the more aggressive the traffic, the more difficult for the agents to achieve convergence, which is also highlighted by the increased standard deviation in all MAB configurations. Again, \gls{tsa} is the best agent, and can converge to the optimal scheme regardless of the value of~$\tau$. Eventually, NA is also able to achieve zero collisions (i.e., $S_{TX}=1$) when $\tau=5$ ms in case of periodic traffic.
	Even the RA approach (our baseline) achieves a successful transmission rate of around $0.9$ as $\tau$ grows, i.e., considering less bandwidth-hungry applications, thanks to the lower collision probability as the contention on the channel becomes less intense.
	Notably, \gls{ucba} is the only method that improves the average accuracy as $\tau$ decreases: the shorter traffic periodicity implies more transmission attempts within the training time, which in turn provides more data to the agent to optimize its decisions. However, oscillations become significant when $\tau = 1.5$ ms.
	\smallskip
	
	\begin{figure}[t!]
		\setlength{\belowcaptionskip}{-0.65cm}
		\centering
		\includegraphics[width=0.83\linewidth]{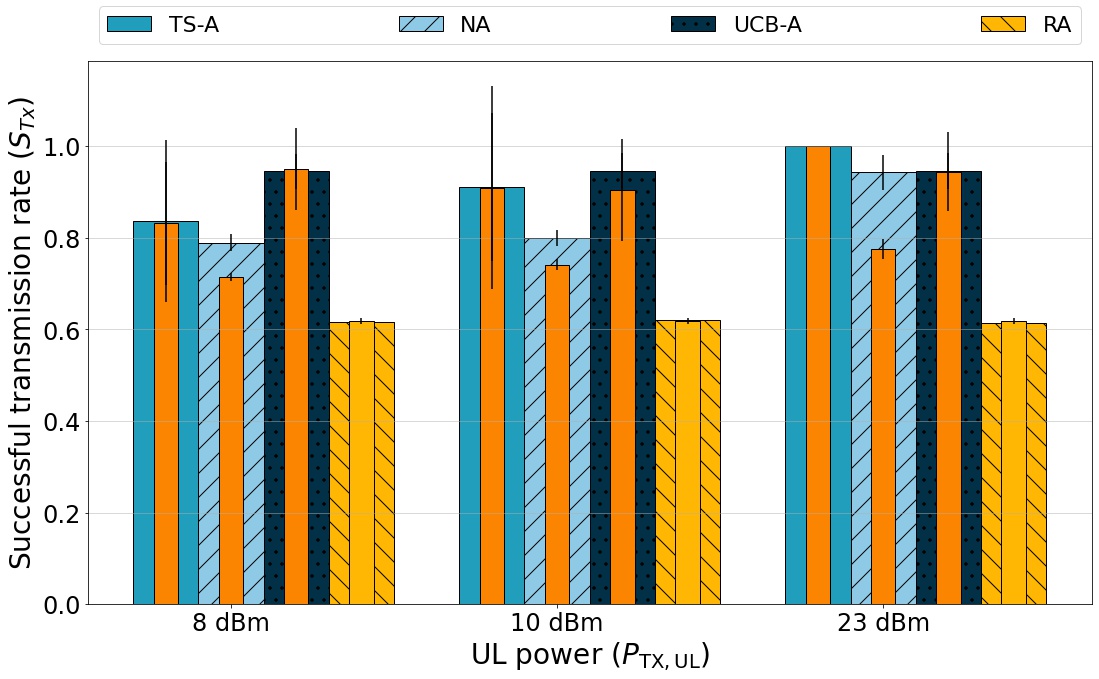}
		\caption{$S_{TX}$ $\pm$ one standard deviation vs. $P_{\rm TX,UL}$ for different MAB agents, after a training time of 60 s, with $N = 100$ and $\tau = 1.5$ ms, with periodic (wide bars) and quasi-periodic (narrow bars) traffic.}
		\label{fig:power_accuracy_after_seconds}
	\end{figure}

	\textbf{Impact of the UL transmission power.} 
	\gls{iiot} devices, such as industrial sensors, may be subject to battery lifetime constraints.
	In light of this, we studied the impact of the UL transmission power $P_{\rm TX,UL} \in \{ 8, 10, 23\}$ dBm on the MAB convergence. While decreasing $P_{\rm TX,UL}$ promotes energy savings and mitigates interference, it may also lead to communication outage when the \gls{sinr} goes below a pre-defined sensitivity threshold, set to $-5$ dB in our simulations.
	In Fig.~\ref{fig:power_accuracy_after_seconds}, with $P_{\rm TX, UL} = 23$ dBm, the outage probability is very small, leading to $S_{TX}\approx 1$ in most configurations (if convergence is achieved).
	As $P_{\rm TX, UL}$ starts decreasing, outage events, besides collisions, lead to additional packet losses, and to a more complex environment. 
	Unlike \gls{tsa} and \gls{na}, \gls{ucba} is less sensitive to this effect. The reasons are twofold. On one side, \gls{na} converges slowly, and is more exposed to retransmissions. At the same time, \gls{tsa} converges quickly to a specific solution, meaning that unpredictable outage events may break the environment statistics underlying the TS algorithm, and lead to unexpected negative feedback from the \gls{gnb}. 
	On the contrary, \gls{ucba} initially explores more, and can better adapt to new configurations in more dynamic scenarios. When $P_{\rm TX, UL}=8$ dBm, \gls{ucba} is the best performing agent, and achieves $+16\%$ $S_{TX}$ compared to \gls{tsa}. 
	\smallskip

	\begin{figure}[t!]
		\setlength{\belowcaptionskip}{0.3cm}
		\centering
		\includegraphics[width=0.83\linewidth]{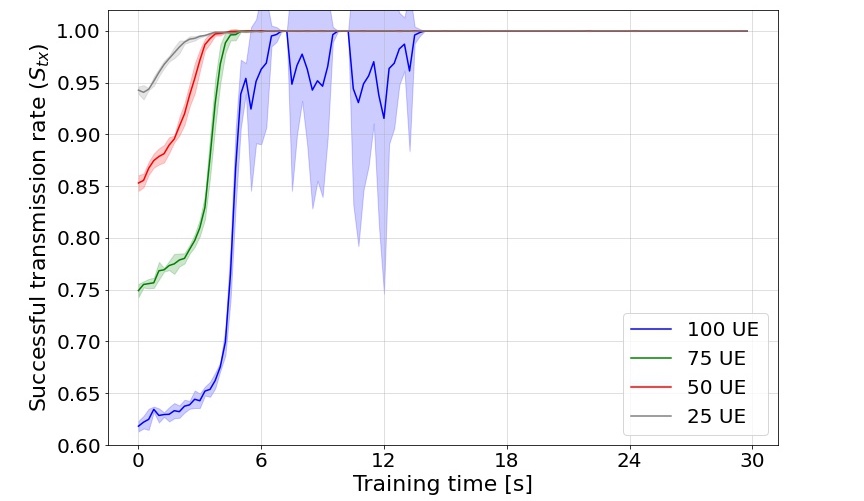}
		\caption{$S_{TX}$ vs. the training time and as a function of $N$, for TS-A with periodic traffic, and $\tau = 1.5$ ms. The curves report mean $\pm$ standard deviation over the simulation runs.\vspace{-0.83cm}}
		\label{fig:acc_vs_time_users_periodic}
	\end{figure}
	\begin{figure}[t!]
		\setlength{\belowcaptionskip}{-0.55cm}
		\centering
		\includegraphics[width=0.83\linewidth]{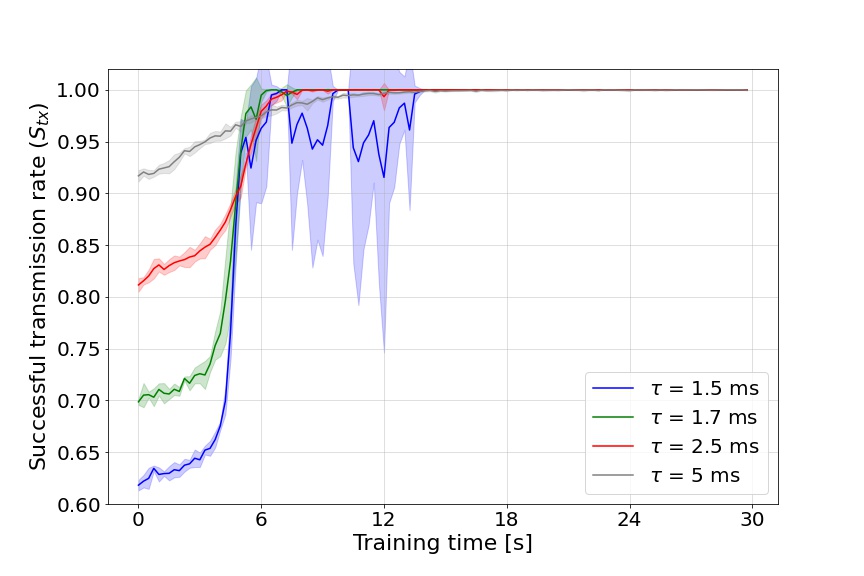}
		\caption{$S_{TX}$ vs. the training time and as a function of $\tau$, for TS-A with periodic traffic, and $N = 100$. The curves report mean $\pm$ standard deviation over the simulation runs.}
		\label{fig:acc_vs_time_period_periodic}
	\end{figure}

	\textbf{TS-A performance.} In view of the above results, we further analyzed TS-A's convergence time to the optimal solution (where no collisions are experienced) as a function of (i) the number of users $N$, and (ii) the traffic periodicity $\tau$.
	In Fig.~\ref{fig:acc_vs_time_users_periodic}, we observe that, as $N$ increases, the TS algorithm takes more time to converge to the best solution, as expected. Notably, the curve with $N = 100$ presents the highest variance, due to the fact that many users are learning an individual policy, leading to a highly non-stationary environment.
	
	In Fig.~\ref{fig:acc_vs_time_period_periodic}, we see that when $\tau$ decreases the convergence time grows accordingly, even though the gap among different configurations is relatively small (convergence is achieved after $\sim8$ s). This is due to the fact that, on the one hand, when the traffic periodicity is short, the problem becomes more complex, as more packets have to be allocated. On the other hand, the agents receive more feedback signals within the same time interval, thus leveraging more data for the training.
	
	
	\subsection{Final remarks}
	Our initial experiments confirm that there exists a MAB configuration for which distributed resource allocation can achieve zero collisions in low latency, i.e., without gNB scheduling grants, thus supporting \gls{urllc}. 
	
	In particular, TS-A is the best performing approach in case of both dense systems and aggressive aperiodic traffic (where conventional semi-persistent/grant-free NR schedulers may fail). \rev{Consequently, our experiments suggest that the Bayesian formulation, together with the exploration strategy of \gls{ts}, are good starting points to build distributed resource allocation in real \gls{iiot} environments, reducing the latency introduced by centralized protocols.} Interestingly, \gls{ucba} works well in complex scenarios, or when UEs communicate with limited power, thus supporting energy efficiency at the expense of some collisions. 
	Moreover, the superior performance in terms of $S_{\rm TX}$ of the \gls{mab} schemes against \gls{ra} shows that machine learning can dramatically reduce, if not completely eliminate, the burden of re-transmissions introduced by 5G-NR-like grant-free access scheduling schemes~\cite{Liu2021}. 
	
	However, distributed resource allocation requires longer training time before convergence, which in real IIoT systems may not be negligible. 
	Still, 
	the training could be run offline, which does not affect the real-time performance of the system (it can be executed when the machine is turned off, e.g., during the calibration of the electro-mechanical processes, or before the service is activated); once active, the service can run rapidly and without significant computational overhead.
	Moreover, our analysis evaluates the training time when the system starts the optimization process from scratch: faster adaptation can be achieved if the system faces limited changes with respect to the initial training scenario, e.g., some components join or leave the system.
	Nevertheless, the trained model still requires retraining when data distributions have deviated significantly from those of the original training set, which involves additional overhead~\cite{kuutti2020survey}. This motivates further explorations in the case of more dynamic systems, that will be carried out as part of our future work.

	\section{Conclusions and Future Work} 
	\label{sec:conclusions_and_future_work}
	
	In this paper we studied the design of user-centric (rather than gNB-centric) distributed (rather than centralized) resource allocation in IIoT scenarios.
	This approach does not involve scheduling grants to be disseminated before UL transmissions, and is thus positioned to better support URLLC compared to conventional scheduling methods. We explored different state-of-the-art MAB agents, for the first time applied to the context of URLLC for IIoT, and identified \gls{tsa} as the best performing implementation, achieving zero collisions in our experiments. \gls{tsa} scales well with the number of users in the system compared to other MAB methods, and still achieves perfect accuracy even considering aperiodic traffic. Notably, \gls{ucba} showed superior performance when the UEs communicate with low power, despite some collision events.
	
	This work opens up new interesting research directions. For example, we will evaluate whether {federated learning}, which optimizes the scheduling policies based on the interaction among the UEs, would result in faster convergence than MAB. 
	
	\section*{Acknowledgment}
	This work has been carried out in the framework of the CNIT National Laboratory WiLab and the WiLab-Huawei Joint Innovation Center.
	
	\bibliographystyle{IEEEtran}
	\bibliography{bibl.bib}

\begin{thebibliography}{10}
\providecommand{\url}[1]{#1}
\csname url@samestyle\endcsname
\providecommand{\newblock}{\relax}
\providecommand{\bibinfo}[2]{#2}
\providecommand{\BIBentrySTDinterwordspacing}{\spaceskip=0pt\relax}
\providecommand{\BIBentryALTinterwordstretchfactor}{4}
\providecommand{\BIBentryALTinterwordspacing}{\spaceskip=\fontdimen2\font plus
\BIBentryALTinterwordstretchfactor\fontdimen3\font minus
  \fontdimen4\font\relax}
\providecommand{\BIBforeignlanguage}[2]{{%
\expandafter\ifx\csname l@#1\endcsname\relax
\typeout{** WARNING: IEEEtran.bst: No hyphenation pattern has been}%
\typeout{** loaded for the language `#1'. Using the pattern for}%
\typeout{** the default language instead.}%
\else
\language=\csname l@#1\endcsname
\fi
#2}}
\providecommand{\BIBdecl}{\relax}
\BIBdecl

\bibitem{giordani2020toward}
M.~Giordani, M.~Polese, M.~Mezzavilla, S.~Rangan, and M.~Zorzi, ``{Toward 6G
  Networks: Use Cases and Technologies},'' \emph{IEEE Communications Magazine},
  vol.~58, no.~3, pp. 55--61, Mar. 2020.

\bibitem{lee2015cyber}
J.~Lee, B.~Bagheri, and H.-A. Kao, ``{A Cyber-Physical Systems architecture for
  Industry 4.0-based manufacturing systems},'' \emph{Manufacturing Letters},
  vol.~3, pp. 18 -- 23, Jan. 2015.

\bibitem{wollschlaeger2017future}
M.~Wollschlaeger, T.~Sauter, and J.~Jasperneite, ``The future of industrial
  communication: Automation networks in the era of the internet of things and
  industry 4.0,'' \emph{IEEE Industrial Electronics Magazine}, vol.~11, no.~1,
  pp. 17--27, Mar. 2017.

\bibitem{cuozzo2022enabling}
G.~Cuozzo, S.~Cavallero, F.~Pase, M.~Giordani, J.~Eichinger, C.~Buratti,
  R.~Verdone, and M.~Zorzi, ``{Enabling URLLC in 5G NR IIoT Networks: A
  Full-Stack End-to-End Analysis},'' in \emph{Joint European Conference on
  Networks and Communications \& 6G Summit (EuCNC/6G Summit)}, 2022.

\bibitem{Yang2021}
P.~Yang, L.~Kong, and G.~Chen, ``Spectrum sharing for {5G/6G} {URLLC}: Research
  frontiers and standards,'' \emph{IEEE Communications Standards Magazine},
  vol.~5, no.~2, pp. 120--125, Apr. 2021.

\bibitem{3gpp.38.321}
3GPP, ``{NR; Medium Access Control (MAC) protocol specification -- Release
  15},'' 3GPP, Technical Specification (TS) 38.321, 2019.

\bibitem{lucas2019capacity}
M.~C. Lucas-Esta{\~n}, J.~Gozalvez, and M.~Sepulcre, ``{On the capacity of 5G
  NR grant-free scheduling with shared radio resources to support
  ultra-reliable and low-latency communications},'' \emph{Sensors}, vol.~19,
  no.~16, p. 3575, Aug. 2019.

\bibitem{boban2021predictive}
M.~Boban, M.~Giordani, and M.~Zorzi, ``{Predictive Quality of Service (PQoS):
  The Next Frontier for Fully Autonomous Systems},'' \emph{IEEE Network},
  vol.~35, no.~6, pp. 104--110, Nov/Dec 2021.

\bibitem{wang2017machine}
M.~Wang, Y.~Cui, X.~Wang, S.~Xiao, and J.~Jiang, ``{Machine learning for
  networking: Workflow, advances and opportunities},'' \emph{IEEE Network},
  vol.~32, no.~2, pp. 92--99, Mar. 2017.

\bibitem{pase2020convergence}
F.~Pase, M.~Giordani, and M.~Zorzi, ``{On the Convergence Time of Federated
  Learning Over Wireless Networks Under Imperfect CSI},'' in \emph{IEEE
  International Conference on Communications Workshops (ICC)}, 2020.

\bibitem{5gaciaiiot}
{5G-ACIA}, ``{5G for Industrial Internet of Things (IIoT): Capabilities,
  Features, and Potential},'' \emph{ZVEI}, Nov. 2021.

\bibitem{intro_mab}
A.~Slivkins, ``{Introduction to Multi-Armed Bandits},'' \emph{Foundations and
  Trends® in Machine Learning}, vol.~12, 2019.

\bibitem{liu2020data}
C.-F. Liu and M.~Bennis, ``{Data-driven predictive scheduling in ultra-reliable
  low-latency industrial IoT: A generative adversarial network approach},'' in
  \emph{IEEE 21st International Workshop on Signal Processing Advances in
  Wireless Communications (SPAWC)}, 2020.

\bibitem{halabian2019distributed}
H.~Halabian, ``{Distributed resource allocation optimization in 5G virtualized
  networks},'' \emph{IEEE Journal on Selected Areas in Communications},
  vol.~37, no.~3, pp. 627--642, Feb. 2019.

\bibitem{hussain2020machine}
F.~Hussain, S.~A. Hassan, R.~Hussain, and E.~Hossain, ``{Machine learning for
  resource management in cellular and IoT networks: Potentials, current
  solutions, and open challenges},'' \emph{IEEE Communications Surveys \&
  Tutorials}, vol.~22, no.~2, pp. 1251--1275, Jan. 2020.

\bibitem{helin2019}
H.~Yang, X.~Xie, and M.~Kadoch, ``Intelligent resource management based on
  reinforcement learning for ultra-reliable and low-latency iov communication
  networks,'' \emph{IEEE Transactions on Vehicular Technology}, vol.~68, no.~5,
  pp. 4157--4169, Jan. 2019.

\bibitem{Le2019}
L.~Liang, H.~Ye, and G.~Y. Li, ``Spectrum sharing in vehicular networks based
  on multi-agent reinforcement learning,'' \emph{IEEE Journal on Selected Areas
  in Communications}, vol.~37, no.~10, pp. 2282--2292, Aug. 2019.

\bibitem{daniel_benji}
D.~Russo and B.~Van~Roy, ``Learning to optimize via posterior sampling,''
  \emph{Mathematics of Operation research}, vol.~39, no.~4, pp. 1221--1243,
  Nov. 2014.

\bibitem{Liu2021}
Y.~Liu, Y.~Deng, M.~Elkashlan, A.~Nallanathan, and G.~K. Karagiannidis,
  ``{Analyzing Grant-Free Access for URLLC Service},'' \emph{IEEE Journal on
  Selected Areas in Communications}, vol.~39, no.~3, pp. 741--755, Mar. 2021.

\bibitem{lattimore_2020}
T.~Lattimore and C.~Szepesvári, \emph{Bandit Algorithms}.\hskip 1em plus 0.5em
  minus 0.4em\relax Cambridge University Press, 2020.

\bibitem{differentible_mab}
C.~Boutilier, C.~wei Hsu, B.~Kveton, M.~Mladenov, C.~Szepesvari, and M.~Zaheer,
  ``{Differentiable Meta-Learning of Bandit Policies},'' in \emph{Advances in
  Neural Information Processing Systems}, 2020.

\bibitem{Bonnefoi2018}
R.~Bonnefoi, L.~Besson, C.~Moy, E.~Kaufmann, and J.~Palicot, ``{Multi-Armed
  Bandit Learning in IoT Networks: Learning Helps Even in Non-stationary
  Settings},'' in \emph{Cognitive Radio Oriented Wireless Networks}, 2018, pp.
  173--185.

\bibitem{polese20193gpp}
M.~Polese, M.~Giordani, and M.~Zorzi, ``{3GPP NR: the cellular standard for 5G
  networks},'' \emph{5G-ITALY White Book}, 2019.

\bibitem{3gpp.38.901}
3GPP, ``{Study on channel model for frequencies from 0.5 to 100 GHz (Release
  16)},'' Technical Specification (TS) 38.901, 2019.

\bibitem{kuutti2020survey}
S.~Kuutti, R.~Bowden, Y.~Jin, P.~Barber, and S.~Fallah, ``{A survey of deep
  learning applications to autonomous vehicle control},'' \emph{IEEE
  Transactions on Intelligent Transportation Systems}, vol.~22, no.~2, pp.
  712--733, Jan. 2020.

\end{thebibliography}

\end{document}